# k-resolved electronic structure of buried heterostructure and impurity systems by soft-X-ray ARPES

V. N. Strocov,[1,][*] L. L. Lev,[1,2,3] M. Kobayashi,[1,4] C. Cancellieri,[1,5] M.-A. Husanu,[1,6] A. Chikina,[1] N. B. M. Schröter,[1] X. Wang,[1] J. A. Krieger[1,7,8] and Z. Salman[7]

[1]Swiss Light Source, Paul Scherrer Institute, 5232 Villigen-PSI, Switzerland
[2]National Research Centre ''Kurchatov Institute'', 123182 Moscow, Russia
[3]Moscow Institute of Physics and Technology, 141700 Dolgoprudny, Russia
[4]Department of Applied Chemistry, University of Tokyo, 113-8656 Tokyo, Japan
[5]Empa, Swiss Federal Laboratories for Materials Science & Technology, 8600 Duebendorf, Switzerland
[6]National Institute of Materials Physics, 077125 Magurele, Romania
[7]Laboratory for Muon Spin Spectroscopy, Paul Scherrer Institute, 5232 Villigen-PSI, Switzerland
[8]Laboratorium für Festkörperphysik, ETH Zürich, CH-8093 Zürich, Switzerland
* Corresponding author (vladimir.strocov@psi.ch)

Angle-resolved photoelectron spectroscopy (ARPES) is the main experimental tool to explore electronic structure of solids resolved in the electron momentum **k**. Soft-X-ray ARPES (SX-ARPES), operating in a photon energy range around 1 keV, benefits from enhanced probing depth compared to the conventional VUV-range ARPES, and elemental/chemical state specificity achieved with resonant photoemission. These advantages make SX-ARPES ideally suited for buried heterostructure and impurity systems, which are at the heart of current and future electronics. These applications are illustrated here with a few pioneering results, including buried quantum-well states in semiconductor and oxide heterostructures, their bosonic coupling critically affecting electron transport, magnetic impurities in diluted magnetic semiconductors and topological materials, etc. High photon flux and detection efficiency are crucial for pushing the SX-ARPES experiment to these most photon-hungry cases.

# Introduction

Angle-resolved photoemission spectroscopy (ARPES) is the main experimental tool to study fundamental electronic structure parameters of crystalline solids – Fermi surface (FS), band structure and one-electron spectral function $A(\omega,\mathbf{k})$ – resolved in electron momentum **k** (Damascelli *et al* 2003, Strocov & Cancellieri 2018). This information is relevant not for the electron system alone but also to its interactions with bosonic degrees of freedom such as phonons and magnons. However, conventional ARPES in the vacuum ultraviolet (VUV) range with photon energies of a few tens of eV is characterized by a photoelectron escape depth $\lambda$ of only a few Å, which hinders its application to buried heterostructures and impurity systems which are at the heart of current electronic and spintronic technologies.

Pioneered by S. Suga's group at SPRing-8 (Suga & Sekiyama 2013), **k**-resolved SX-ARPES operates in the photon energy range of a few hundred eV, which brings a few fundamental advantages compared to VUV-ARPES (Fadley 2012, Strocov *et al* 2014, Suga & Tusche 2015):

(1) High photoelectron kinetic energies increase $\lambda$ by a factor of 3-5 as described, in a first approximation, by the "universal curve" of the inelastic mean free path of electrons in solids (Powell *et al* 1999). With the probing depth being of the order of $3\lambda$, this increase is already sufficient to access buried interfaces;

(2) For three-dimensional (3D) materials, the increase of $\lambda$ has an added value that it reduces intrinsic broadening of the out-of-plane momentum $k_z$ defined, by the Heisenberg uncertainty principle, as $\Delta k_z = \lambda^{-1}$ (Strocov 2003). In combination with the free-electron dispersion of high-energy final states, the resulting sharp definition of $k_z$ allows precise determination of the electronic structure in 3D. Starting from paradigm metals like Al (Hofmann *et al* 2002) and W (Papp *et al* 2011, Medjanik *et al* 2016), this has been demonstrated by the pioneering studies of 3D nesting of the Fermi surface (FS) in chalcogenides (Strocov *et al* 2012, Weber *et al* 2018); dimensionality and correlation effects in high-$T_c$ pnictides (Derondeau *et al* 2017), cuprates (Suga *et al* 2004, Matt *et al*. 2018, Horio *et al*. 2018) and Mott materials (Xu *et al* 2014); electron-phonon interaction in cuprates (Tsunekawa *et al* 2008) and its band dependence in manganates extending beyond the Migdal approximation (Husanu *et al* 2019-1), structure stabilization mechanisms in AlNiCo quasicrystals (Rogalev *et al* 2015); FS nesting in ruthenates possibly connected with superconductivity (Sekiyama *et al* 2004); effects of lattice distortions on colossal magnetoresistance in manganites (Lev *et al* 2015); dimensionality-driven metal-insulator transitions in iridates (Schütz *et al* 2017); origin of ferromagnetism in $CrO_2$ related to its 3D electronic structure measured through the $Cr_2O_3$ surface layer (Bisti *et al* 2017); bulk Rashba spin splitting in the non-centrosymmetric BiTeI (Landolt *et al* 2012) and multiferroic Mn-doped $\alpha$-GeTe (Krempaský *et al* 2016); *f*-electron hybridization in heavy-fermion materials (Yano *et al* 2008, Höppner *et al*. 2013, Chen *et al*. 2018), verification of the topological fermions in Weyl semimetals (Lv *et al* 2015, Xu *et al* 2015, Xu *et al* 2017, Lv *et al* 2017, Di Sante *et al* 2017, Xu *et al* 2018, Yao *et al* 2019) and chiral topological semimetals (Schröter *et al* 2019); mixed dimensionality in topological insulators (Manzoni *et al* 2016); and many more. More examples of SX-ARPES applications to 3D materials can be found in previous reviews (Suga & Sekiyama 2013, Strocov *et al* 2014, Suga & Tusche 2015). However exciting, these applications will stay beyond the main scope of this review, which is focused on the more challenging cases of buried systems;

(3) The SX-ARPES energy range covers *L*-edges of the transition metals and *M*-edges of the rare-earth ones. Resonant photoemission (ResPE) at these edges can be used to achieve elemental and even chemical state specificity (Molodtsov 1997, Olsson *et al* 1996, Kobayashi *et al* 2014, Chikina *et al* 2018). In combination with an increased probing depth, this virtue of SX-ARPES allows the experimenter to lock the ARPES response to particular interfaces or impurity atoms in buried systems.

We note however that a significant resonant enhancement of coherent spectral intensity can in general be expected only in systems where the participating conduction and valence band states are sufficiently localized and have the same orbital character. In this case they have sufficient overlap with the localized core level as well as with each other in the ResPE process (Chikina *et al* 2018). This situation typically realizes in transition metals with open *d*-shell (Sc through Ni) and rare earths with open *f*-shell (La through Lu). As these elements are typically used as magnetic dopants, ResPE is particularly suited for studies of magnetism.

The main challenge of SX-ARPES include, however:

(1) A dramatic loss of photoexcitation cross-section of valence states, typically by a few orders of magnitude compared to the UV energy range (Yeh & Lindau 1985);

(2) The photoelectron wavelength in the soft-X-ray energy range becomes comparable with the thermal atomic motion amplitudes, which relaxes the dipole selection rules and reduces the coherent spectral structure (Braun *et al* 2013). Apart from the thermal effects, the coherent ARPES intensity suffers from intrinsic phonon excitation due to electron-phonon interaction (Hofmann *et al* 2002) which can be seen as a recoil (Suga & Sekiyama 2009);

(3) On the instrumental side, gradually relaxing with an increase of excitation energy are the energy resolution, roughly proportional to $h\nu$, and angular resolution proportional to $\sqrt{E_k}$, where $E_k$ is the photoelectron kinetic energy.

These difficulties have until recently severely limited the usefulness of **k**-resolved SX-ARPES. A breakthrough in practical applications of this technique has been demonstrated at the ADRESS beamline of the Swiss Light Source. As we will see in the examples below, a combination of high photon flux of above $10^{13}$ photons/s/0.01%BW delivered by the beamline (Strocov *et al* 2010) with the grazing-incidence endstation geometry (Strocov 2013, Strocov *et al* 2014) have not only overcome the cross-section problem, but also allowed pushing SX-ARPES to the most photon-hungry cases of buried heterostructures and impurities where the ARPES signal is attenuated by scattering in the overlayers.

We note that pushing ARPES into the multi-keV energy range of hard-X-ray photon (HX-ARPES) further extends $\lambda$ 50-100 Å and more (Powell *et al* 1999, Fadley 2012). Successful **k**-resolved HX-ARPES has recently been reported for some materials like W and GaAs (Gray *et al* 2011, Gray *et al* 2012). However, this technique suffers from further dramatic reduction the VB cross-section and loss of the coherent spectral weight, in particular due to recoil which smears **k** and energy definition (Suga & Sekiyama 2009).

Before unfolding specific examples of **k**-resolved SX-ARPES, we will illustrate the ability of this technique to access buried systems with a basic example of GaAs(100) grown by MBE and protected from atmospheric contamination by an amorphous As capping layer with a thickness of ~10 Å. Fig. 1 shows experimental ARPES intensity images corresponding to the ΓKX line of the bulk Brillouin zone (BZ) which were acquired at increasing $h\nu$ values. The low-energy image is dominated by a structureless signal from the amorphous As overlayer. With the increase of *hv*, the ARPES images develop astonishingly clear dispersive structures formed by coherent photoelectrons coming from the GaAs underlayer. We immediately identify the canonical manifold of the light-hole (LH), heavy-hole (HH) and spin-orbit split (SO) bands of GaAs which are now seen through the thick capping layer. This observation unambiguously identifies the increase of $\lambda$ coming with an increase of *hv* into the SX-ARPES energy range. Further details of this educative experiment, including linear dichroism of the ARPES response of the LH and SO vs HH bands, can be found in Kobayashi *et al* 2012. We will now turn to applications of SX-ARPES to a few specific heterostructure and impurity systems.

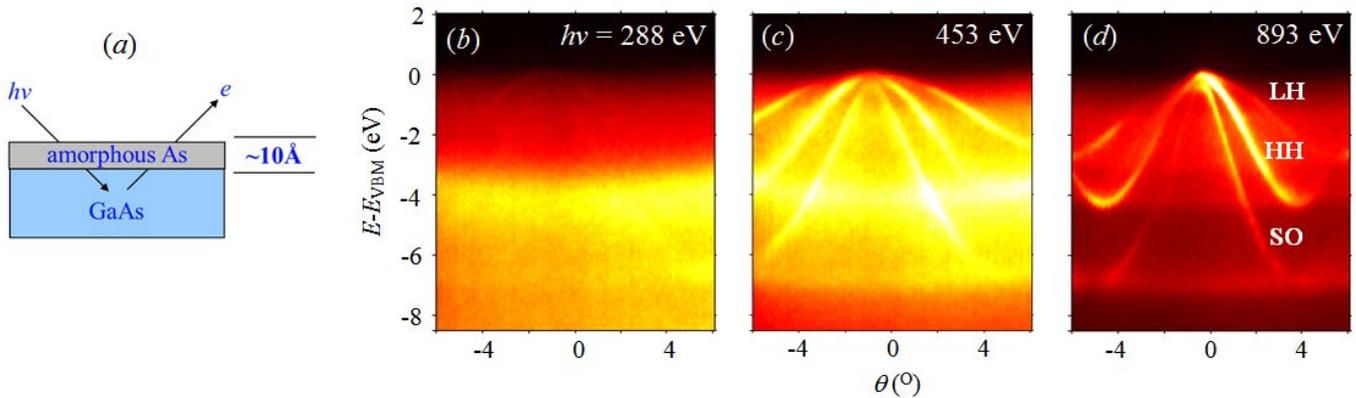

Fig. 1. SX-ARPES experiment on As-capped GaAs(100): (*a*) Sketch of the sample; (*b-d*) ARPES images along the $\overline{\Gamma}$KX line measured at the indicated *hv* values. Development of the GaAs band structure signal with *hv* evidences the increase of $\lambda$. Adapted from Kobayashi *et al* 2012.

# Buried heterostructures

The large probing depth of SX-ARPES illustrated above is crucial to access buried interfaces and heterostructures, which are the workhorse of present electronic devices.

## Semiconductor heterostructures: GaAlN/GaN high-electron-mobility transistors

A recent example of the use of SX-ARPES to study semiconductor technology is our investigation of GaAlN/GaN high-electron-mobility transistor (HEMT) heterostructures (Lev *et al* 2018). A large conduction band (CB) offset augmented by strong electric polarization forms a deep potential well at the interface that confines a highly mobile 2D electron gas on the GaN side (Medjdoub & Iniewski 2015). This well typically embeds 1-2 quantum well (QW) states, as sketched in Fig. 2(*a*). The spatial separation of these QW states from the defect-rich GaAlN barrier layer allows the electrons to escape defect scattering and thereby dramatically increases their mobility. This idea is the fundamental operating principle of the HEMTs, which are finding their applications in a wide range of microwave devices, such as cell phones and radars.

Because of a relatively large thickness of the GaAlN layer of ~3 nm, shown in Fig. 2(*b*), the buried QW states can only be accessed in the *hv* range above ~1000 eV. Fig. 2(*c*) reproduces the experimental in-plane Fermi surface (FS) of the buried QW states measured at *hv* = 1057 eV. The FS appears as tiny electron pockets located at the $\overline{\Gamma}$-points of the 2D interfacial BZ. Their intensity distribution over different BZs reflects the 2D Fourier composition of the QW wavefunctions. A momentum-distribution curve (MDC) at Fermi energy $E_F$ measured across the $\overline{\Gamma}_{11}$-point, shown in Fig. 2(d), resolves Fermi momenta $k_F$ of the two QW states. Remarkably, we discover a significant planar anisotropy of the electronic structure along $\overline{\Gamma M}$ and $\overline{\Gamma K}$ from the difference of the corresponding $k_F$ values of ~11%. In the experimental band dispersions in Fig. 2(e) this translates into a difference of the effective mass $m^*$ of ~22%. Caused by a piezoelectrically active relaxation of atomic positions near the GaN/AlN interface, this anisotropy leaves the linear (low-field) transport properties isotropic, but manifests itself in non-linear (high-field) electron transport as anisotropy of the saturation drift velocity and current. These findings bear direct technological implications for an improvement of the high-power performance of the GaAlN/GaN HEMT devices by aligning conductive channels along optimal momentum directions.

Fig. 2(f) reproduces the experimental FS measured in the ($k_x$,$k_z$) coordinates under variation of $hv$. The 2D character of these states manifests itself by the straight FS contours without any $k_z$ dispersion. However, their ARPES response shows periodic intensity oscillations peaked where $k_z$ hits the out-of-plane reciprocal vectors (Strocov 2018). This shows that the QW wavefunctions are derived from the conduction band minimum of bulk GaN at the Γ-point. In this way, the SX-ARPES experiment not only fully characterizes the FS and band dispersions of the QW states, but also their wavefunctions. Such identification can be essential, for example, for heterostructures based on layered materials that are relevant for valleytronics applications (for entries see Schaibley et al 2016) where the CB can include several local minima that are almost degenerate in energy, but separated in **k**. The demonstrated direct **k**-space imaging of the fundamental electronic structure characteristics –

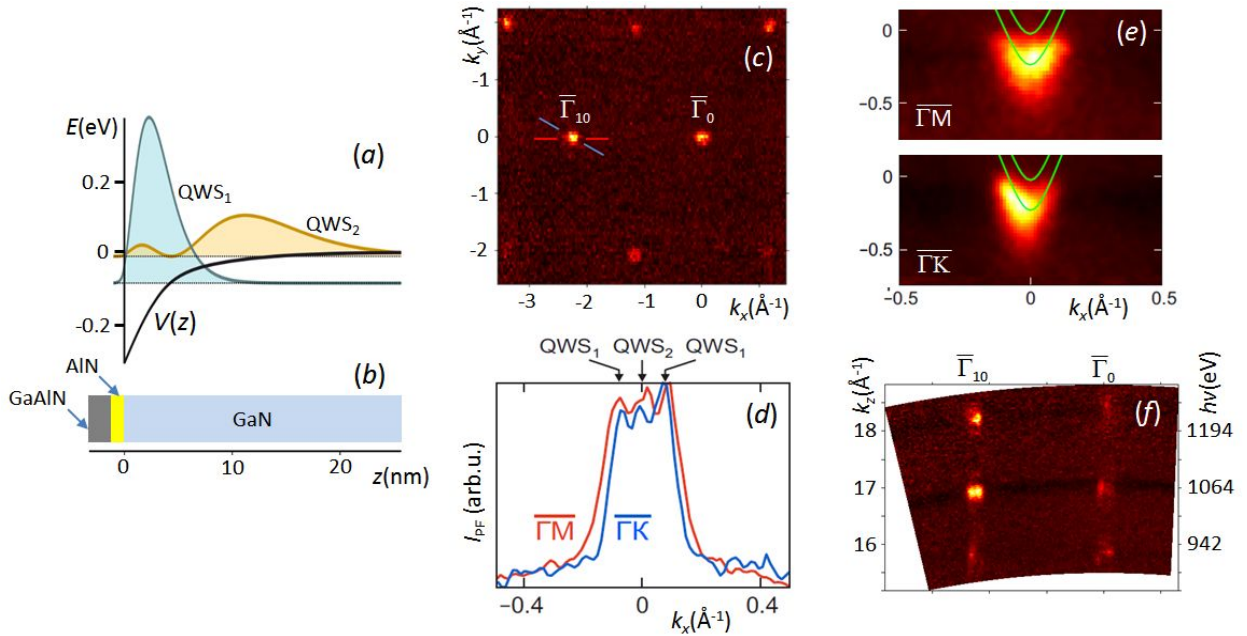

**Fig. 2.** ARPES data on the buried QW states in AlGaN/GaN heterostructure: (a) Envelope function of the two QW states embedded at (b) AlGaN/GaN heterostructure; (c) Experimental in-plane FS measured at $hv$ = 1057 eV; (d) MDCs at $E_F$ along the $\overline{\Gamma M}$ and $\overline{\Gamma K}$ azimuths (marked in c). The MDC peaks identify $k_F$ of the two QW states. The evident electronic structure anisotropy translates into anisotropic non-linear transport properties; (e) Experimental band dispersions of the QW states along $\overline{\Gamma M}$ and $\overline{\Gamma K}$; (f) Experimental FS cross-section in the $\overline{\Gamma M}$ azimuth as a function of $k_z$. The ARPES response blows up where the out-of-plane $k_z$ coincides with the Γ-points, evidencing that the QW state wavefunctions are derived from the CBM of bulk GaN (adapted from Lev et al 2018).

FS and its Luttinger volume, band dispersions and occupancy, Fourier composition of wavefunctions – of GaAlN/GaN HEMT heterostructures makes a quantitative step in understanding the physics of these devices compared to conventional optical and magnetotransport experiments.

Semiconductor heterostructures are presently one of the most active areas of SX-ARPES research. The studied systems include the $SiO_2$/SiC interfaces (Woerle *et al* 2017), doped $\delta$-layers in Si, interfaces between superconductors and semiconductors with large spin-orbit splitting such as Al/InAs conceived as 2D prototypes of Majorana fermion systems for quantum computing, etc.

## Oxide heterostructures: $LaAlO_3$/$SrTiO_3$

Interfaces and heterostructures of strongly correlated transition metal oxides (TMOs) can give rise to new physical phenomena which cannot be anticipated from the properties of individual constituents, with the new functionalities having great promise for future device applications (Mannhart & Schlom 2010). One of such systems is the $LaAlO_3$/$SrTiO_3$ (LAO/STO) interface which embeds a mobile 2D electron system (2DES) whose unusual properties include field effect, coexisting ferromagnetism and superconductivity, etc. The 2DES is localized at the STO side of the interface and is, according to theory, formed by the Ti $t_{2g}$-states which are split into the in-plane $d_{xy}$-derived and out-of-plane $d_{xz/yz}$-derived bands, Fig. 3 (*a*) (Cancellieri *et al* 2014). Resonant SX-ARPES is ideally suited to penetrate through the LAO layer with a thickness of a few u.c. and access the interfacial electrons (Koitzsch *et al* 2013, Berner *et al* 2013).

The experimental angle-integrated ResPE spectrum from the LAO/STO interface across the Ti $L_3$- and $L_1$-edges, shown in Fig. 3(*b*), pinpoints the 2DES signal resonating around $h\nu$ = 460 eV and 466 eV (Cancellieri *et al* 2013; 2016). We note that these resonances of the $t_{2g}$-derived 2DES are shifted from the unoccupied $Ti^{4+}$ $t_{2g}$-states towards the $e_g$-ones at higher $h\nu$, that is explained by **k**-conservation between the conduction- and valence-band states coupled in the ResPE process (Chikina *et al* 2018). The FS map (*c*) acquired at the $L_3$-resonance with circularly polarized incident X-rays immediately reveals the circular $d_{xy}$- and elliptical $d_{xz/yz}$-derived FS sheets (Berner *et al* 2013, Cancellieri *et al* 2014). Their intensity variations across the BZs reflects the Fourier composition of the corresponding wavefunctions. We note that the Luttinger count of the experimental FS seen by SX-ARPES significantly exceeds the carrier concentration $n_s$ ~ $10^{13}$ $e$/cm$^2$ observed in transport experiments, identifying an electronic phase separation (EPS) at the interface where micro- and nanoscale conducting 2DES areas coexist with insulating ones (Strocov *et al* 2019). This fact is crucial for the coexistence of superconductivity and weak ferromagnetism in LAO/STO (Scopigno *et al* 2016, Ariando *et al* 2011). The EPS can be tuned by variation of the oxygen deficiency during the LAO/STO growth or under X-ray irradiation (Strocov *et al* 2018, 2019).

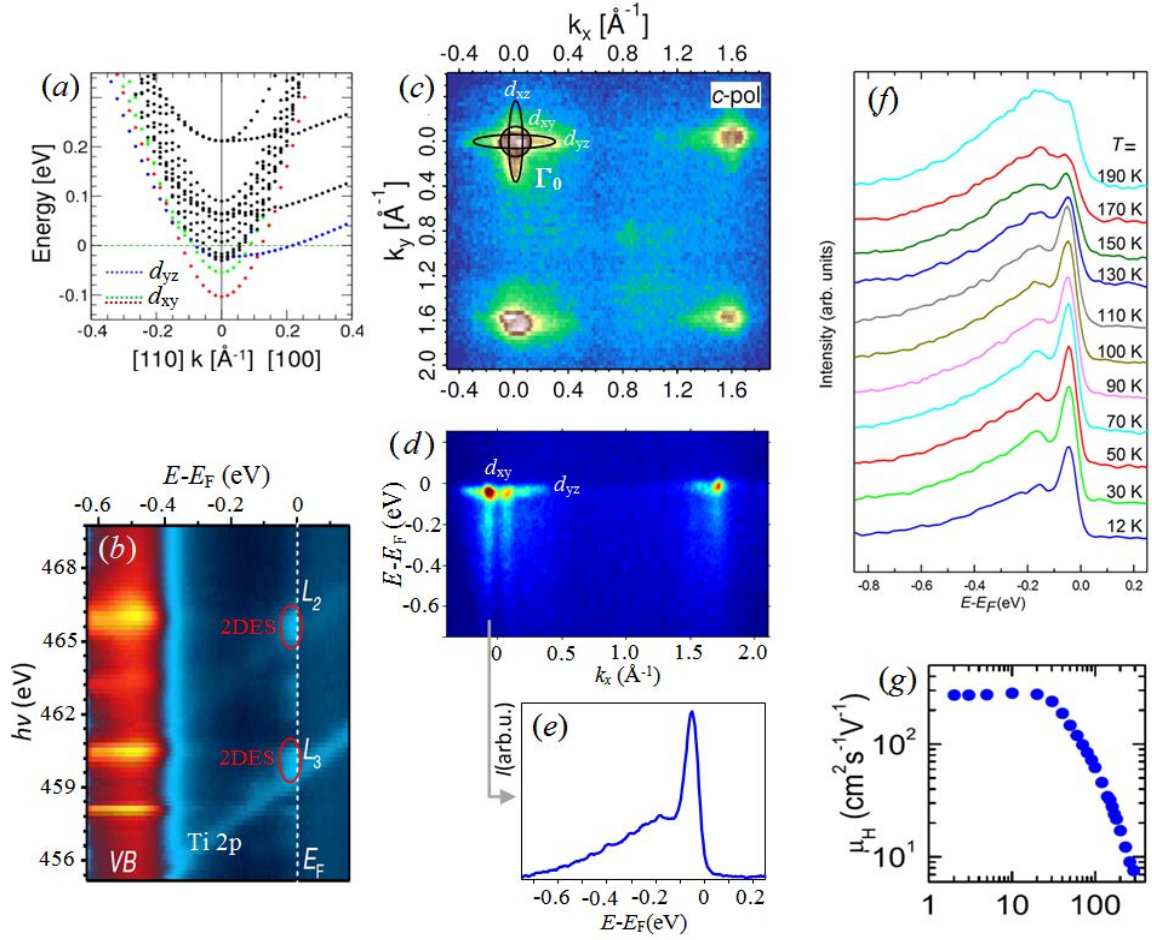

**Fig. 3.** SX-ARPES of the LAO/STO interface: (*a*) Theoretical band structure formed by the $d_{xy}$- and $d_{xz/yz}$-derived bands; (*b*) Angle-integrated ResPE intensity through the Ti 2*p* edge, showing up the 2DES signal at $E_F$; (*c*) FS map formed by the $d_{xy}$- and $d_{xz/yz}$-derived FS sheets; (*d*) Experimental $E(\mathbf{k})$ of $d_{xy}$- and $d_{xz/yz}$-derived bands; (*e*) Experimental $A(\omega,\mathbf{k})$ whose peak-dip-hump lineshape identifies polaronic nature of the charge carriers; (*f*) Temperature dependence of the lineshape. The polaronic coupling quenches the quasiparticle peak, explaining (*g*) the drop of electron mobility around 200K. Adapted from Cancellieri *et al* 2016.

Finally, pushing our SX-ARPES setup to its resolution limit of around 35 meV resolves the individual bands, Fig. 3(*c*), where the *s*-polarization selects the antisymmetric $d_{xy}$- and $d_{yz}$-derived states (Cancellieri *et al* 2014; 2016). We note, strikingly, two prominent waterfalls, where the corresponding energy-distribution curves of ARPES intensity $I(E_B,\mathbf{k})$ (*e*), representing the one-electron spectral function $A(\omega,\mathbf{k})$, have a peak-dip-hump lineshape characteristic of strong electron coupling to bosonic degrees of freedom. In our case the electrons couple to the high-energy LO3 breathing phonon mode in STO (Cancellieri *et al* 2016, Strocov *et al* 2018) to form large polarons (Moser *et al* 2013, Mishchenko *et al* 2011). This increases the interface charge carrier's $m^*$ by a factor ~2.5, which fundamentally limits the 2DES mobility. Doping of the interface with oxygen vacancies increases $n_s$ and concomitantly reduces the polaronic coupling strength. Temperature dependence of the $A(\omega,\mathbf{k})$ lineshape (*f*) shows that the polaronic coupling scales up with temperature and finally quenches the quasiparticle peak, explaining the mysterious mobility drop (*g*) above 200K observed in transport properties of LAO/STO (Gariglio *et al* 2009). This phenomenon is caused by electron coupling to the low-energy TO1 polar phonon mode which changes its frequency from ~18 to ~14 meV across the tetragonal to cubic phase transition in STO at 105K. The

2DES at the LAO/STO interface is thus formed by large polarons involving at least two phonons with different energy and thermal activity. We note that this high-resolution SX-ARPES experiment has been the first observation of polaronic effects at buried interfaces in the most direct way as manifested by the one-electron $A(\omega,\mathbf{k})$. Such an experiment on an interface buried below a ~17-Å thick overlayer and probed with energy resolution of 35 meV is presently the SX-ARPES instrumentation forefront. A comprehensive overview of SX-ARPES studies on the LAO/STO interface can be found in (Strocov *et al* 2018).

Further examples of oxide interfaces include dimensionality-tuned electronic structure of epitaxial $LaNiO_3/LaAlO_3$ superlattices (Berner *et al* 2015), the 2DES formed in the $GdTiO_3$/STO multilayer system (Nemšák *et al* 2016), multiferroic $BaTiO_3/La_{1-x}Sr_xMnO_3$ interfaces (Husanu *et al* 2019-2), etc.

## Hybrid heterostructures: $SiO_2$/EuO/Si spin injectors

A recent example relevant for spintronics is the spin injector heterostructure $SiO_2$/EuO/Si (Lev *et al* 2016). EuO is one of the most promising routes for spintronics, for a review see (Caspers *et al* 2015). This material delivers almost 100% spin-polarized electrons which are relevant for spin-filter applications and can be integrated into the wide-spread Si technology.

Our experiments on $SiO_2$/EuO/Si heterostructures, which are illustrated in Fig. 4(*a*), used photon energies above 1 keV, allowing access to the Si substrate. The experimental spectrum shown in Fig. 4 (*b*) was measured with *hv* = 1120 eV, which tunes **k** to the valence band maximum (VBM) of bulk Si. On top of the dispersionless $Eu^{2+}$ state, we recognize the textbook manifold of the light-hole and heavy-hole bands of bulk Si. The energy difference between the upper edge of the $Eu^{2+}$ multiplet and the VBM of Si directly measures the EuO/Si band offset as 0.8 eV, justifying the spin injecting functionality of this interface. The observation of the Si bands through the 30-Å thick $SiO_2$/EuO layer is an impressive example of the penetrating power of SX-ARPES.

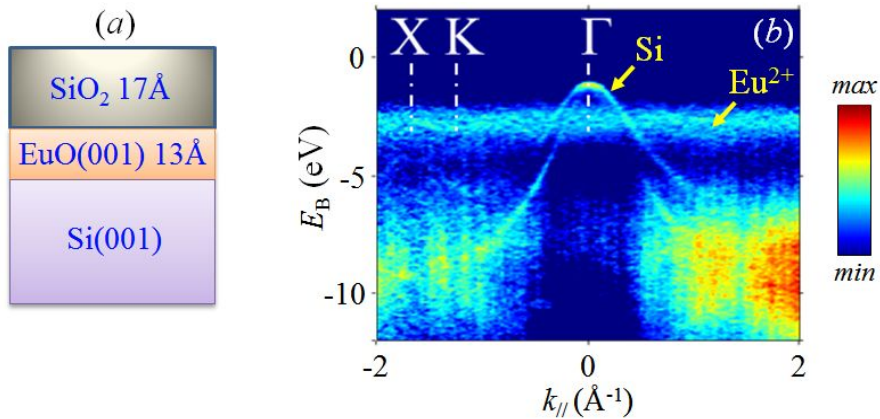

**Fig. 4.** $SiO_2$/EuO/Si spin injector: (*a*) Scheme of the sample; (*b*) SX-ARPES image (angle-integrated component subtracted) measured at *hv* = 1120 eV, which reveals the multiplet of the $Eu^{2+}$ *f*-states on top of the three-dimensional $E(\mathbf{k})$ of bulk Si along the ΓKX direction. The observed band offset of 0.8 eV justifies the spin injecting functionality of the EuO/Si interface. Adapted from Lev *et al* 2017.

Another interesting example is an interface between amorphous Si and STO (Chikina *et al* 2019). Under X-ray radiation, oxygen atoms diffuse from the STO substrate to Si, and the $t_{2g}$ electrons released by the oxygen vacancies left behind in STO create there conducting regions.

# Impurity systems

Access to the local electronic structure of impurity atoms with an atomic concentration on the percent and even per mille level is a sort of a 'needle in a haystack' problem which can be solved by leveraging the elemental and chemical specificity of resonant SX-ARPES.

## Semiconductors: Mn-doped GaAs

The power of resonant SX-ARPES can be illustrated by our study on the paradigm magnetic semiconductor (Ga,Mn)As (Kobayashi *et al* 2014) where the Mn doping induces ferromagnetism (FM) associated with hole carriers. One model of the FM in (Ga,Mn)As (for entries see Ohya *et al* 2011) is the *p-d* exchange, where the Mn 3*d* states form an impurity band (IB) above $E_F$ and the FM is induced by its exchange interaction with the valence-band holes. Another is the double-exchange model, where the IB is placed at $E_F$ and the FM is stabilized by hopping of spin-polarized holes within the IB. Understanding of the exact physics of (Ga,Mn)As requires knowledge of the IB energy position and its hybridization with the GaAs host states.

Fig. 5 reproduces the results by Kobayashi *et al* 2014 measured on (Ga,Mn)As samples with a Mn concentration of 2.5%. In the $L_3$-edge X-ray absorption spectroscopy (XAS) spectrum (*a*), the first peak at 640 eV is related to the FM substitutional Mn atoms and the second at 644 eV to the paramagnetic (PM) interstitial ones. The corresponding resonant ARPES images (*b*) were measured with *s*-polarized incident X-ray, selecting the HH-bands of the host GaAs. When $h\nu$ is tuned on the FM-peak of the XAS spectrum, the ARPES image immediately reveals the Mn 3*d* derived IB, whose energy position just below $E_F$ identifies it as the principal state injecting the FM charge carriers in (Ga,Mn)As. Furthermore, measurements with *p*-polarized X-rays (see the original work by Kobayashi *et al* 2014) show an intensity enhancement of the LH-band of GaAs at this excitation energy which identifies its hybridization with the Mn-derived IB. These results, combining the previous double-exchange and *p-d* exchange models (Ohya *et al* 2011), are summarized in the band diagram in Fig. 5(*c*). The dispersionless character of the IB suggests that the GaMnAs physics should be understood starting from the Anderson impurity model embedding hybridization of the Mn *d*-orbitals with the GaAs band electrons. These resonant SX-ARPES results of Kobayashi *et al* 2014 are consistent with previous HX-ARPES results (Gray *et al* 2012) as well as with a later HX-ARPES study where excitation of Bragg-reflection standing X-ray waves was used to decompose the electronic structure into element- and momentum-resolved components (Nemšák *et al* 2018). From a device perspective, the most important result of these spectroscopic studies is that most of the FM charge carriers in (Ga,Mn)As are derived from the IB and thus have small mobility. This fact prohibits high-frequency applications of (Ga,Mn)As. Recently we also applied this methodology to another diluted magnetic semiconductor, the n-doped (In,Fe)As (Kobayashi *et al* 2019), where sharp dispersions of the FM electron carriers have identified their mobility much exceeding that of the FM holes in GaMnAs.

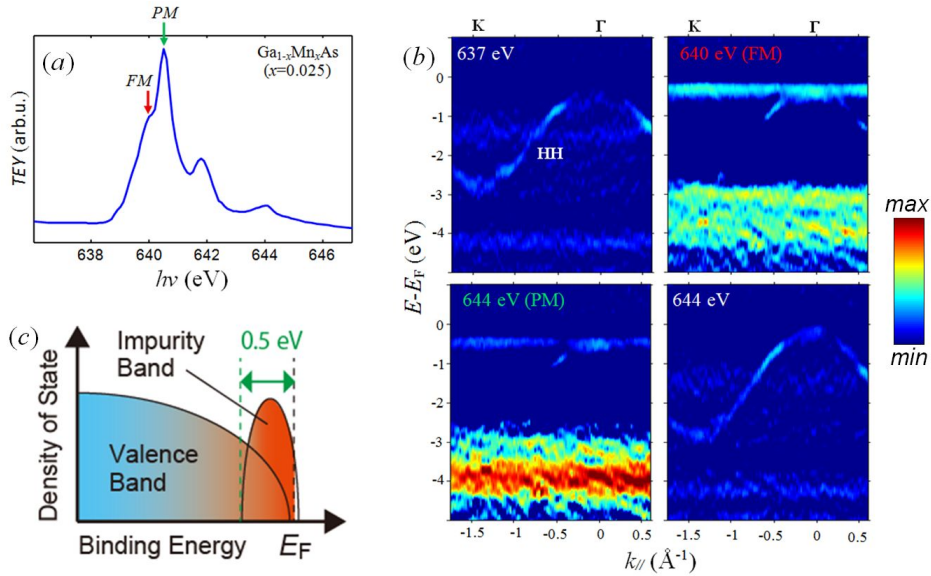

**Fig. 5:** SX-ARPES of Mn-doped GaAs: (*a*) Mn $L_3$ XAS spectrum showing the FM- and PM-components; (*b*) A series of ARPES images (represented in the second derivative -d$^2$I/dE$^2$ > 0) taken across the resonance. Tuning $h\nu$ onto the FM peak unleashes the Mn-derived IB in the vicinity of $E_F$ injecting the FM charge carriers; (*c*) Established band diagram. Adapted from Kobayashi *et al* 2014.

### Topological materials: V-doped (Bi,Sb)$_2$Te$_3$

Another example is magnetic V-impurities in the topological insulator (Bi,Sb)$_2$Te$_3$. Such impurities can induce FM, opening a gap in the Dirac point of the topological surface states otherwise protected by the time-reversal symmetry. This can allow the realization of the quantum anomalous Hall (QAH) effect, featuring quantized dissipationless edge-state transport in the absence of an external magnetic field (Chang et al 2013). Currently this effect is limited to temperatures far below the Curie temperature. Possible reasons for this include magnetic inhomogeneities (Lachman *et al* 2015) and bulk/impurity states contributions to the transport (Peixoto *et al* 2016). Therefore, it is important to understand the effect of the V impurities onto the electronic structure.

Fig. 6 compiles the experimental results of Krieger *et al* (2017) for (Bi,Sb)$_2$Te$_3$ doped with 6% of V, for which the QAH was previously reported (Chang *et al* 2015). (*a*) shows the V *L*-edge XAS spectrum, and (*b*) the difference between the on- and off-resonance ARPES intensity at the excitation energies marked in the XAS spectrum as $h\nu_B$ and $h\nu_A$. This differential image reflects essentially the V-weight in the valence-band wavefunctions. We note dispersions of the (Bi,Sb)$_2$Te$_3$ host states hybridized with V and, most importantly, a flat V-derived IB at $E_F$ where the Dirac cones should form the gap responsible for the QAH-effect. These experimental results are reproduced by DFT-based Coherent Potential Approximation (CPA) calculations of the V-weight shown in (*c*). In principle, the large density of states associated with the IB would have hindered the realization of the QAH in V-doped (Bi,Sb)$_2$Te$_3$. A viable conjecture to resolve this discrepancy invokes non-uniform magnetism in this material demonstrated by muon spin rotation (μSR) experiments, which reveal a magnetic inhomogeneity and partial magnetic volume fraction. In this case V-rich FM islands impose magnetic field on the largely V-depleted PM host, where the IB is suppressed, allowing thereby the QAH. An extensive account of this combined ResPE and μSR study can be found in the original work (Krieger *et al* 2017). Our findings for the V-doped (Bi,Sb)$_2$Te$_3$ are in line with the Mn 3*p* resonant ARPES study of Mn-doped Bi$_2$Se$_3$ which has also found a significant dopant weight near the Dirac point (Sánchez-Barriga *et al* 2016) and evidenced remarkable structure modifications induced by the incorporation of Mn atoms (Rienks *et al* 2018).

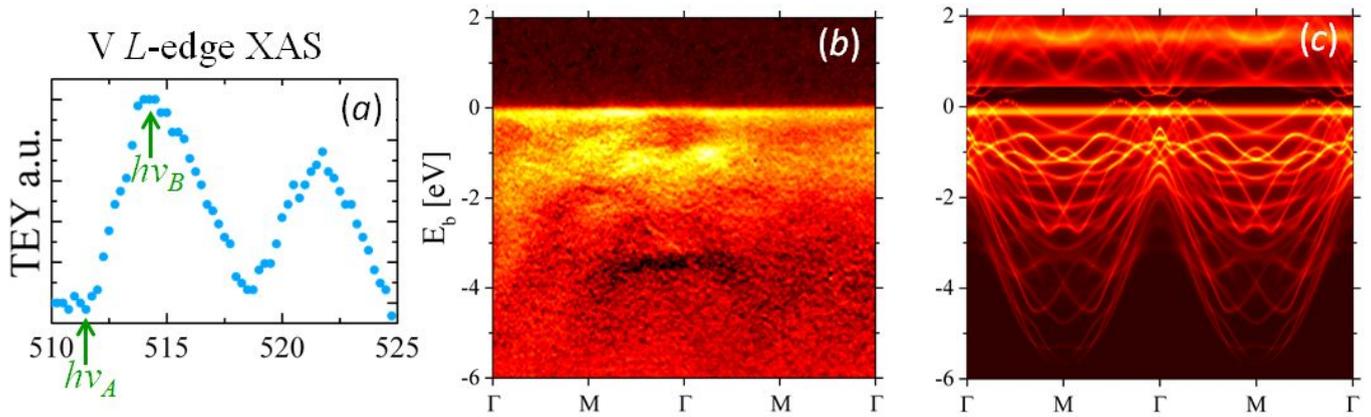

**Fig. 6:** SX-ARPES of V-doped topological $(Bi,Sb)_2Te_3$: (*a*) V *L*-edge XAS spectrum; (*b*) Differential ARPES image between the on- and off-resonance *hv* marked in (*a*) as $hv_B$ and $hv_A$, respectively; (*c*) DFT-CPA calculations of the V spectral weight. The V-derived IB band at $E_F$ counteracts the QAH-effect. Adapted from Krieger *et al* 2017.

Other applications of SX-ARPES to impurity systems include the multiferroic (Ge,Mn)Te where Mn doping of the host $\alpha$-GeTe breaks the time-reversal protected degeneracy of the Rashba-split bands, inducing a Zeeman gap whose magnitude scales with the Mn concentration (Krempaský *et al* 2016).

# Outlook

The spectroscopic power of SX-ARPES arises from merging its **k**-resolution with enhanced probing depth delivered by high-energy photoelectrons and chemical specificity achieved with resonant photoexcitation. These virtues make SX-ARPES ideal to access the electronic structure of buried heterostructure and impurity systems which are in the heart of electronic, spintronics, and quantum devices. These applications of SX-ARPES have been illustrated here with a few selected scientific cases, including buried quantum-well states in GaAlN/GaN high-electron-mobility transistors, band offsets in $SiO_2$/EuO/Si spin injectors, polaronic charge carriers at the buried $LaAlO_3/SrTiO_3$ interface, local electronic structure of magnetic impurities in the semiconducting Mn-doped GaAs and topological V-doped $(Bi,Sb)_2Te_3$, etc. High photon flux and grazing-incidence experimental geometry are must-haves for SX-ARPES experiments in such photon-hungry cases.

Further advance in spectroscopic abilities of SX-ARPES implies increasing its detection efficiency, allowing access to yet deeper buried heterostructures and smaller impurity concentrations, and sharpening its energy resolution towards yet finer energy scales. The latter is required to study low-energy bosonic excitations such as phonons and magnons, playing a crucial role in optical and transport properties of devices, or order parameters such as superconducting gaps in high-$T_c$ materials (Chatterjee *et al* 2010). Extension of SX-ARPES experiments to in-situ field effect will allow electronic structure studies on operando quantum devices. Another vector of SX-ARPES development is its spin resolution. The necessary spin detection efficiency increase can be achieved with multichannel spin detectors based, for example, on imaging spin filters utilizing spin-dependent low-energy electron reflection (Schönhense *et al* 2015, Tusche *et al* 2015, Medjanik *et al* 2016). Another multichannel concept is the imaging Mott (iMott) spin analyzer utilizes Mott scattering of high energy (~40 keV) electrons to achieve spin resolution and imaging-type electron optics to enable energy- and angle-multichannel detection (Strocov et al 2015). The spin-resolving capabilities will push the SX-ARPES experiment into the totally unexplored domain of spin textures of buried heterostructure and impurity systems, which is of paramount importance for novel spintronics devices.

## Acknowledgements

We thank S. Suga, J. Minár, A. S. Mishchenko, J.-H. Dil, M. Shi, R. Claessen, M. Sing, A. Fujimori, D. Feng, A. X. Gray, C. S. Fadley, J.-M. Triscone, F. Lechermann, P. Blaha, A. Filippetti, E. E. Krasovskii, E. Chulkov, W. Drube, P. Willmott, T. Schmitt, G. Aeppli and many other condensed-matter experts for valuable discussions. M.-A.H. was supported by the Swiss Excellence Scholarship grant ESKAS-no. 2015.0257, A.C. by the Swiss National Science Foundation under grant No. 200021_165529 and J.A.K. under grant No. 200021_165910. The research of M.K. at Swiss Light Source was supported by a grant from the Japan Society for the Promotion of Science.